\documentclass{jfm}
\usepackage{graphicx}
\usepackage{newtxtext}
\usepackage{newtxmath}
\usepackage{natbib}
\usepackage{hyperref}
\hypersetup{
    colorlinks = true,
    urlcolor   = blue,
    citecolor  = black,
}

\newcommand{\RomanNumeralCaps}[1]
\linenumbers

\shortauthor{S. Sharma,N. K. Chandra, A. Kumar, S. Basu}
\title{Shock induced  atomisation of a liquid metal droplet}

\author{Shubham Sharma\aff{1}, Navin Kumar Chandra\aff{1}
Aloke Kumar\aff{1},
  \and   Saptarshi Basu\aff{1,2}
\corresp{\email{sbasu@iisc.ac.in}}
 }

\affiliation{

\aff{1}Department of Mechanical Engineering, Indian Institute of Science, Bangalore-560012, India

\aff{2}Interdisciplinary Centre for Energy Research, Indian Institute of Science, Bangalore-560012, India}

\begin{document}
\maketitle

\begin{abstract}
The present study uses Galinstan as a test fluid to investigate the shock-induced aerobreakup of a liquid metal droplet in a high Weber number regime $(We \sim 400 - 8000)$. Atomization dynamics is examined for three test environments: oxidizing (Galinstan-air), inert (Galinstan-nitrogen), and conventional fluids (DI water-air). Due to the readily oxidizing nature of liquid metals, their atomization in an industrial scale system is generally carried in inert atmosphere conditions. However, no previous study has considered gas-induced secondary atomization of liquid metals in inert conditions. Due to experimental challenges associated with molten metals, laboratory scale models are generally tested for conventional fluids like DI water, liquid fuels, etc. The translation of results obtained from conventional fluid to liquid metal atomization is rarely explored. Here a direct multi-scale spatial and temporal comparison is provided between the atomization dynamics of conventional fluid and liquid metals under oxidizing and inert conditions.  The liquid metal droplet undergoes breakup through Shear-Induced Entrainment (SIE) mode for the studied range of Weber number values. The prevailing mechanism is explained based on the relative dominance of droplet deformation and KH wave formation. The study provides quantitative and qualitative similarities for the three test cases and explains the differences in morphology of fragmenting secondary droplets in the oxidizing test case (Galinstan-air) due to rapid oxidation of the fragmenting ligaments. A phenomenological framework is postulated for predicting the morphology of secondary droplets. The formation of flake-like secondary droplets in the Galinstan air test case is based on the oxidation rate of liquid metals and the properties of the oxide layer formed on the atomizing ligament surface.

\end{abstract}

\begin{keywords}
Authors should not enter keywords on the manuscript, as these must be chosen by the author during the online submission process and will then be added during the typesetting process (see \href{https://www.cambridge.org/core/journals/journal-of-fluid-mechanics/information/list-of-keywords}{Keyword PDF} for the full list).  Other classifications will be added at the same time.
\end{keywords}

{\bf MSC Codes }  {\it(Optional)} Please enter your MSC Codes here

\section{Introduction}
\label{sec:Introduction}

Droplet aerobreakup is a phenomenon where a primary droplet formed during bulk liquid atomisation is further fragmented into multiple secondary droplets due to disruptive forces, such as aerodynamic and shear forces. These disruptive forces aim to disintegrate the droplet while surface tension, viscosity, and elasticity resist fragmentation. If the disruptive forces overcome the restoring forces, the droplet deforms, and an imbalance between the two forces can lead to secondary atomization. This process is significant in both natural and industrial operations, including the formation of raindrops \citep{villermaux2009single} and various industrial applications such as fuel atomization in combustion engines, spray drying, and pesticide spraying \citep{Odenthal,Rajamanickam_2017,mondal2019spray,sharma2021dynamics}.  As a result, extensive experimental and numerical works are carried out to uncover the complicated mechanisms and identify the different parameters governing the interaction phenomenon.

\hspace{0.2cm} \indent Here, we study the aerobreakup of a droplet through a shock droplet interaction mechanism. The shock drop interaction process consists of two stages: the initial shock interaction (stage I) and droplet breakup dynamics (stage II, occurring at later times) \citep{sharma2021shock}. Many experimental and numerical studies have investigated the droplet breakup dynamics using the shock tube method. However, limited research has been conducted on the significance of the early wave dynamics in the breakup analysis \citep{sharma2021shock,Sembian_2016, Meng_and_Colonius_2015, Tanno2003, sun2005unsteady}. \cite{Sembian_2016} used experimental and numerical techniques to analyze the evolution dynamics of reflected, transmitted, and diffracted waves during the interaction of an incident shock wave with a cylindrical water column. They observed the possibility of cavitation in the droplet due to the expansion wave focusing at higher shock Mach numbers. \cite{Meng_and_Colonius_2015} and  \cite{Guan_2018} conducted numerical simulations to illustrate the development of recirculating flow near the droplet's pole region and an upstream jet at the droplet wake, which eventually assists in droplet deformation and breakup. These  studies suggest the importance of considering the early-stage shock wave interaction in analyzing droplet breakup dynamics during a shock drop interaction process.

\hspace{0.2cm} \indent Post interaction with a shock wave,  shock-induced gas flow interacts with the liquid droplet, which majorly governs the atomization dynamics \citep{sharma2021shock}. According to \cite{Hinze_1955}, the breakup mode of a droplet is dependent on the  Weber number $\left(We= \frac{\rho_gV_i^2D_o}{\sigma}\right)$ and Ohnesorge number $\left(Oh=\frac{\mu_l}{\sqrt{\rho_l \sigma D_o}}\right)$, which was further extended by \cite{KRZECZKOWSKI_1980} to identify the transition points on the $We-Oh$ regime. Here, $\mu_l$ and $\rho_l$ are the dynamic viscosity and density of the droplet, $\sigma$ is the surface tension of the air-liquid interface, $D_o$ is the initial droplet diameter before shock interaction, $\rho_g$ and $V_i$ are the density and velocity of gas flow at post-shock conditions. The Weber number accounts for the relative dominance of aerodynamic/inertial force to the surface tension force, while the Ohnesorge number accommodates the effect of liquid viscosity on breakup dynamics. The fragmentation modes and corresponding breakup transition criteria have been consolidated by numerous reviews \citep{sharma2022advances,PILCH_1987,Guildenbecher2009,Theofanous_ARF_2011}. Based on different flow Weber number values, primarily five modes of aerodynamic breakup of a droplet have been obtained for $Oh<0.1$, i.e., bag breakup ($24>We>11$), bag-stamen/plume breakup ($65>We>24$), multi-bag breakup ($85>We>65$), sheet thinning/stripping breakup ($350>We>85$), and catastrophic breakup ($We>350$) \citep{Jain2015,Guildenbecher2009}. Based on hydrodynamics instabilities responsible for such breakups, \cite{Theofanous2008,Theofanous_2012} reclassified these breakup modes. For $We<100$, the previously identified bag, bag-stamen, and multi-bag breakup modes were consolidated as Rayleigh-Taylor Piercing (RTP) mode (based on Rayleigh-Taylor Instability). Similarly, at higher Weber numbers $(We>1000)$, the previously identified sheet thinning/stripping and catastrophic modes were consolidated as Shear-Induced Entertainment (SIE) mode (based on Kelvin Helmholtz instability). The present work uses the same description for the two modes.

\hspace{0.2cm} \indent The literature on droplet aerobreakup is primarily based on the aerobreakup 
 of conventional test fluids such as DI water, oils, liquid fuels, etc., which differ significantly from liquid metals in various aspects. Aerobreakup of liquid metal has not received much attention in the literature. Still, it has important applications in metal powder production, thermal spray coatings, explosive detonations, metalized propellant combustion, and liquid metal cooling systems. Several challenges exist in studying the atomization of liquid metals, including the applicability of conventional fluid results to liquid metals, the marginally explored area of the liquid metal breakup, and the contrasting fluid properties of liquid metals compared to conventional fluids, such as higher density, higher surface tension, lower kinematic viscosity, and high oxidation rates. To avoid the influence of temperature-dependent fluid properties variation, liquid metal experiments are generally carried out on pure metals/alloys that are in the liquid state at room temperature, such as Galinstan (Gallium 68.5\%
, Indium 21.5\%
 and tin 10\%
), EGaIn, or Mercury. Galinstan, a Gallium-based eutectic liquid metal alloy, is popular due to its non-toxicity. However, Galinstan is readily oxidized in environmental conditions where oxygen concentration is $>$ 1ppm \citep{liu2011characterization}. The exact values of its fluid properties are debatable in the literature. Surface tension values are reported in the range of 500-700 mN/m \citep{liu2011characterization,arienti2019comparison,plevachuk2014thermophysical,handschuh2021critical}. This range is primarily due to the presence of oxidizing conditions in which measurements were undertaken. On the other hand, the viscosity and density values of $\sim2$ mPa.s and $\sim6440$ $Kg/m^3$, respectively, are consistent among different studies \citep{plevachuk2014thermophysical,xu2012effect,handschuh2021critical}. The rapid oxidation of galinstan alloy forms a thin elastic layer (size $\sim O(1)$nm \citep{jia2019liquid}) of gallium oxide which impacts the fluid's rheological properties \citep{dickey2008eutectic,xu2012effect,elton2020dramatic}. Due to the challenges associated with liquid metals, only a few studies have been reported on their secondary atomization.

\hspace{0.2cm} \indent  \cite{hsiang1995drop} conducted preliminary work on the secondary atomization of mercury droplets, where the bag breakup mode was found for $We \sim10$. Similar to the observations which were made for other conventional fluids. \cite{chen2018galinstan} investigated the galinstan liquid column breakup by shock-induced crossflow for Weber number values ranging up to 250 and compared the results with those for water. They found that the breakup behavior for the two fluids was very similar, with similar values of Weber number for mode transition. However, differences were observed in the shape of fragments, with irregular-shaped daughter droplets formed in the case of the Galinstan column due to the oxide layer formed on the exposed surface of droplets.  Similar observations were made through numerical simulation \citep{arienti2019comparison}. \cite{hopfes2021secondary,hopfes2021experimental} studied the atomization of galinstan droplets for moderate Weber numbers ($We = 10-104$) and found similar breakup modes and critical Weber number values for regime transition for Galinstan and conventional fluids with $Oh<0.1$. However, the bag formation occurred with lesser inflation, and its rupture appeared to be quicker for Galinstan droplets due to the oxide layer formation on the droplet surface. The role of oxidation was further verified by comparing the atomization results of Galinstan droplet with Field's metal. The Field's metal has a lower oxidation rate and therefore has higher bag inflation and higher breakup time.

 \hspace{0.2cm} \indent The present work is focused on a high Weber number regime ($We \sim O(1000)$), which itself comes with several associated challenges. The aerodynamic breakup of a liquid droplet is a high-speed phenomenon where the droplet disintegration is completed within the time scales of $O(100)$ $\mu$s, particularly at higher Weber numbers. High-speed interaction dynamics is difficult to perceive through numerical and experimental means due to the short time and length scales  of the breakup phenomenon. The disintegration of a millimeter size primary droplet into micron-sized daughter droplets within a time scale of a few microseconds requires high spatial-temporal resolution. The usage of high exposure times (even within microsecond order) results in the motion blur of fast-moving atomised droplets, which can lead to a misinterpretation of the actual phenomenon (as discussed by \citet{Theofanous2008}). Therefore, a highly sophisticated and optimised experimental arrangement is required for effectively capturing the multi-scale nature of the droplet's aerodynamic breakup and involved physical phenomenon. Although several recent experimental studies have attempted to investigate some of these aspects \citep{Theofanous2008, Biasiori-Poulanges_2019, jackiw_ashgriz_2021}, a comprehensive study providing benchmark measurement data for numerical simulation, which covers a wide parametric range and the complete evolution dynamics of the interaction phenomenon, is still lacking. In our previous works, we attempted to fill this gap by investigating high-Weber number atomization of Newtonian and viscoelastic droplets using high-quality experimentation \citep{sharma2021shock, chandra2022shock}. In this study, we extend our efforts to explore the atomization dynamics of a liquid metal droplet.

\hspace{0.2cm} \indent The discourse above indicates that the characteristics of oxide-forming liquid metals during fragmentation differ from those of traditional fluids, and a complete comprehension of their atomization dynamics is yet to be achieved. Laboratory scale analysis for liquid metal atomization is generally conducted using conventional fluids as test liquids. Therefore, a thorough understanding of the applicability of conventional fluid results to liquid metals can help to bridge the gap between conventional fluids and hot molten melts and improve industrial processes involving liquid metal atomization. Existing research on liquid metal droplets has mainly concentrated on low to moderate Weber numbers $(We < 100)$ interactions and compared the results of an oxide-forming liquid metal droplet with conventional fluids that lack oxidation characteristics. Here, it must be noted that due to the rapid oxidation rates of liquid metals, their industrial scale atomization is carried in inert atmosphere conditions \citep{mandal2022experimental}. Particularly for the application involving metal powder production for additive manufacturing applications, liquid jets generally interact with high-speed supersonic gas jets. The flow Mach numbers of 4-6 are achievable, which can easily lead to a high local Weber number ($> O(1000)$) for the primary droplets, which remain unexplored in the literature. The present study addresses these issues, investigates the shock-induced atomization dynamics of a Galinstan droplet in a high-Weber number regime $(We \sim 1000$ to $8000)$, and conducts experiments in an inert atmosphere to prevent Galinstan oxidation. The results can be directly compared to conventional fluids, unlike the approach followed in the literature. The comparison between three test environments, namely conventional fluids (water-air), oxidizing fluids (Galinstan-air), and inert fluids (Galinstan-nitrogen), are made at multiple spatial and temporal scales to provide a qualitative and quantitative comparison of three test cases. Finally, a phenomenological framework is provided for predicting the morphology of secondary droplets. The experimental shock tube setup is simplified by using an exploding wire-based technique to generate the required shock wave. Its inherent decaying flow characteristics provide a more realistic approach to studying the atomization phenomenon than the conventional shock tube setup. The advanced and optimized experimental arrangement used in this study addresses the concern of high spatial-temporal imaging of interaction phenomena, providing a suitable benchmark for future numerical validations. 

\hspace{0.2cm} \indent This paper is organised as follows; \S\ref{sec:Materials and methods} provides the details of the experimental setup and methodology used.  Results and discussions are provided in \S\ref{sec:Results and discussion}. Details are provided on global observation of the phenomenon (\S\ref{subsec:Global view of atomization dynamics}), governing mechanism(\S\ref{subsec:Mechanism of Galinstan droplet breakup in SIE mode}), qualitative (\S\ref{subsec:Atomization dynamics of Galinstan droplet at high Weber number}) and quantitative (\S\ref{subsec:Quantitative estimation of droplet deformation and KH waves magnitude}) comparison of three test cases and phenomenological framework for predicting morphology of the secondary droplets (\S\ref{subsec:Phenomenological framework for secondary droplets morphology}). The conclusions are provided in \S\ref{sec:Conclusions}.

\section{Materials and methods}
\label{sec:Materials and methods}

\subsection*{Exploding-wire based shock tube}
\label{subsec:Exploding wire based shock tube}

\begin{figure}%
\centering
\includegraphics[width=1\textwidth]{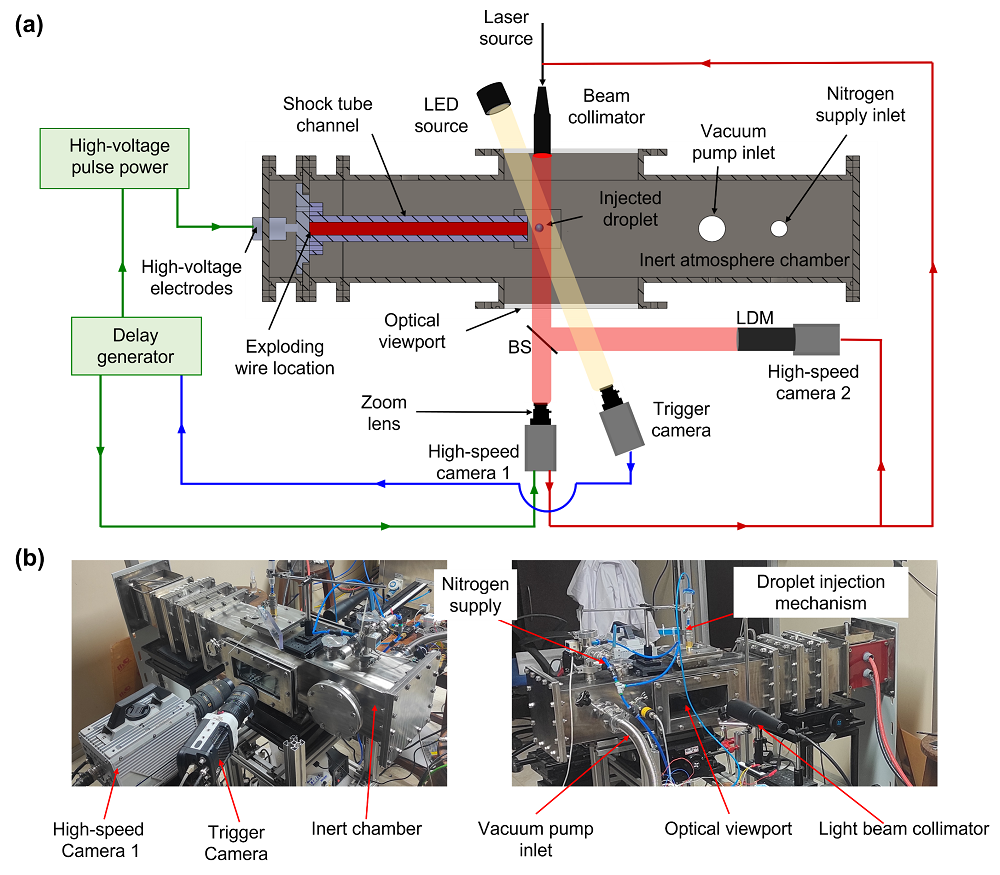}
\caption{Exploding wire based shock tube setup incorporated with inert atmosphere chamber. (a) Schematic diagram showing experimental setup, electrical wiring diagram, and high-speed imaging system. (b) Photograph of the experimental arrangements.}
\label{Ch4:fig:Experimental setup}
\end{figure}

\begin{figure}
\centering
\includegraphics[width=1\textwidth]{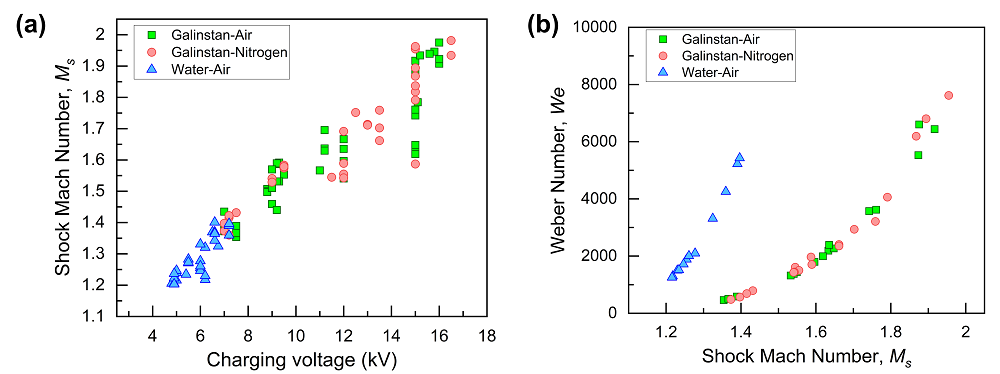}
\caption{Range of Non-dimensional numbers. (a) Variation of shock Mach number ($M_s$) with the capacitor charging voltage of a pulse power system, (b) Variation of Weber number($We$) with shock Mach number ($M_s$). Similar Weber number values are maintained for comparing Galinstan-air, Galinstan-nitrogen, and water-air test cases.}
\label{Ch4:fig:Weber number regime}
\end{figure}

Figure \ref{Ch4:fig:Experimental setup} depicts an exploding-wire (EW) based shock tube setup, which generates a shock wave that interacts with a freely falling droplet. The shock tube operates by passing a high-voltage pulse (in the order of kilo-volts and microseconds duration) through a thin metallic wire (35 SWG, bare copper wire) placed between two high-voltage electrodes, resulting in rapid Joule heating, melting, and vaporization of the wire into a column of dense vapours. The formed vapour column expands and generates a cylindrical blast wave, which is transformed into a normal shock by the rectangular confinement of the shock tube flow channel (320 mm $\times$ 40 mm $\times$ 20 mm). A 2 kJ pulse power system (Zeonics Systech, India Z/46/12) provides a high-voltage pulse by discharging a 5 $\mu$F capacitor across the exploding wire. The charging voltage is varied from 5 $kV$ to 15 $kV$, producing shock waves with varying strengths (see figure \ref{Ch4:fig:Weber number regime}a). The shock Mach number $(M_s=U_s/v)$ ranges from 1.2 to 2.0, leading to a broad range of Weber number variation ($\sim$ 400 - 8000), as shown in figure \ref{Ch4:fig:Weber number regime}b. Here, $U_s$ represents the shock speed at the instant of shock interaction with the droplet. It is measured using the distance moved by the shock wave in two consecutive camera frames, while $v$ refers to the speed of sound in the medium ahead of the shock wave, i.e., gas at 0.35 bar and 298 K in the current experiment. Weber number in the present work is defined as:
\begin{equation}
    \label{eq: Weber number}
    We = \frac{Disrupting pressure}{Laplace pressure} = \frac{2(P_o-P_1)}{\sigma/D_o}
\end{equation}

Here $P_o$ and $P_1$ are the stagnation pressure and static pressure at the droplet equator (i.e., at the windward point) and pole, respectively.  The values of $P_o$ and $P_1$ are estimated using compressible flow theory for shock-induced airflow at the time instant of shock droplet interaction \citep{anderson1990modern}. The above definition of $We$ is consistent for both incompressible and compressible flows.
The importance of representing Weber number in this form is explained in supplementary figure S1.

\indent \hspace{0.2 cm} Several references provide a detailed overview of the exploding-wire (EW) technique and its application in shock-wave generation \citep{chandra2022shock,sharma2021shock,Sembian_2016,liverts2015mitigation}. Compared to conventional diaphragm-based shock tubes, the EW technique offers advantages such as smaller test facilities, easier operation, a wide range of shock Mach numbers (from 1 to 6 \citep{Sembian_2016}), and high repeatability between tests. However, it is worth noting that conventional shock tube setups provide a uniform flow for a longer duration (around $10^0-10^2$ ms) compared to the droplet breakup timescales (around $10^1-10^3$ $\mu$s). On the other hand, the inherent properties of blast-wave-based shock tube setups cause fluid properties like gas velocity, density, and pressure to decay rapidly with time. Therefore, it is not possible to make a direct comparison of droplet aerobreakup achieved by these two different shock generation techniques. Blast-wave-based setups provide a parallel approach for investigating shock interactions, as shown in several studies \citep{chandra2022shock,ram2012implementation, igra2013review,pontalier2018experimental,supponen2018jetting}. However, their utilization for studying droplet aerobreakup is a new approach \citep{sharma2021shock,chandra2022shock, Sembian_2016}. Therefore, discussing the transient aspects of shock and shock-induced flow properties associated with the present setup is important. Such features of the current experimental setup were discussed in previous work performed on the same setup \citep{chandra2022shock,sharma2021shock}. The decaying flow characteristics are more practical scenarios, and atomization results of liquid metal droplets in such flow conditions can have direct implications in applications like closed coupled atomizers, detonation engines, etc.  Many practical scenarios can benefit more from the blast wave-based shock tube setups, compared to the ideal conditions from conventional shock tube setups \citep{pontalier2018experimental, igra2013review}. The current study's primary objective is to compare the atomization dynamics of a liquid metal droplet with a DI water droplet, which is not affected by the choice of experimental setup as both liquids are tested under the same conditions.

\color{black}

\subsection*{Imaging setup}
\label{sec:Imaging setup}

The droplet aerobreakup process is captured using high-speed shadowgraphy imaging (see figure \ref{Ch4:fig:Experimental setup}). Specifically, we utilized a Photron SA5 camera synchronized with a Cavitar Cavilux smart UHS laser to freeze the interaction phenomenon in the 10-40 ns time scale, preventing the streaking of high-speed secondary droplets that could lead to observational errors. Side-view images are obtained at different zoom settings to capture different aspects of the droplet breakup. The light from a high-speed laser is transformed into a parallel beam using a beam collimator (Thorlabs, BE20M-A) to uniformly illuminate the camera's field of view. The droplet is detected by an image based trigger camera (PHANTOM Miro 110, coupled with 100 mm macro-lens Tokina) as it fell into the field of view of the high-speed camera-1 (Photron SA5, coupled with a Sigma DG 105 mm), resulting in the projection of its shadow on the camera-1 sensor. A trigger signal generated by the trigger camera is sent to a digital delay generator (BNC 575), which precisely provides delayed trigger signals to a high-voltage pulse power system and high-speed camera-1, allowing us to capture the atomization dynamics of the fragmenting droplet in the camera's field of view and at a desired location. Another high-speed camera-2 coupled with a long-distance microscope (LDM, Questar QM100) is used for simultaneous high-zoom imaging. High-speed camera-2 and the laser are operated as slaves to high-speed camera-1.

\indent \hspace{0.2 cm} To obtain a global view of the droplet morphology during the breakup process, a zoomed-out imaging technique is employed. High-speed camera-1 is used for this purpose, capturing the interaction dynamics at 40000 frames per second with a frame size of $640 \times 264$ pixels. The resulting pixel resolution is 52.6 $\mu$m/pixel, providing a field of view of 33.6 mm $\times$ 13.91 mm. Further, for a more detailed view of the evolution of Kelvin-Helmholtz (KH) waves on the droplet surface, a zoomed-in imaging technique is used.  A long-distance microscope coupled with a high-speed camera 2, captures the growth of KH waves at an imaging rate of 75000 fps. A frame size of 256 $\times$ 312 pixels is used, resulting in a spatial resolution of 13.16 $\mu$m/pixel and a field of view of 3.37 mm $\times$ 4.11 mm. The gas flow direction is from left to right in all the experimental images presented in this article.

\subsection*{Sample preparation and characterization}

The Galinstan alloy used in this study is procured from Parmanu Dhatu Nigam, India. The liquid metal is utilized directly in the experiments conducted in an air environment. However, in inert atmosphere studies, it is crucial to eliminate the pre-existing metal oxide from the bulk liquid due to Galinstan's tendency to oxidize readily. This is achieved by utilizing the chemical reduction method, which involves injecting the liquid metal into a 5M HCl solution \citep{kim2013recovery,handschuh2021critical}. The solution reduces the oxidized Galinstan through the following reaction.

\begin{equation}
Ga_2O_3 (\textit{s}) + 6HCl (\textit{l}) \to 2GaCl_3 (\textit{s}) + 3H_2O (\textit{l})
    \label{eq:Galinstan reaction}
\end{equation}

While loading onto the droplet injection unit, the reduced galinstan is poured into the injection tube, which is partially filled with 5M HCl solution to prevent air contact. This ensures that the galinstan alloy remains completely unoxidized during the injection process. On the other hand, DI water can be directly added to the injection unit without any special precautions. The droplets of the test fluid are injected into the test chamber using pressurized air pulses. To perform this injection, the test fluid is held in a plastic tube connected to a stainless steel leur-lock needle (size 20 G) through a ball valve. To create an inert atmosphere inside the chamber, the ball valve is closed (see figure \ref{Ch4:fig:Experimental setup}). However, after maintaining the chamber conditions and during the injection of a droplet, it is necessary to open the valve. The top end of the plastic tube is connected to a high-pressure air line through an electronically controlled solenoid valve, which exerts a pressure pulse on the test fluid by controlling the opening of the solenoid valve.

\indent \hspace{0.2 cm} Table \ref{Table_1} displays the properties of the three test fluids. The surface tension values of DI water and Galinstan are determined using the pendant drop method using \textit{ImageJ} software. On the other hand, the data for viscosity and density are obtained from a previous study \cite{hopfes2021experimental}.

\begin{table}
  \begin{center}
\def~{\hphantom{0}}
  \begin{tabular}{cccc}
      \textbf{Property}  & \textbf{Water} &   \textbf{Galinstan-Air} & \textbf{Galinstan-Nitrogen}\\[6pt]
        Density $(Kg/m^3)$ & 997  & 6440 & 6440 \\[6pt]
        Viscosity $(mPa.s)$ & 1  & 2.4 & 2.4 \\[6pt]
        Surface tension $(mN/m)$ & 72  & 665 & 548 \\[6pt]
  \end{tabular}
  \caption{Properties of the test liquids. Surface tension is measured using the pendant drop method. Other properties are taken from \cite{hopfes2021experimental}}
  \label{Table_1}
  \end{center}
\end{table}

\subsection*{Inert atmosphere test facility}
The schematic and actual photograph of the inert atmosphere test facility is presented in figure \ref{Ch4:fig:Experimental setup}. The facility has a stainless steel chamber (measuring $15 \times 15 \times 75$ $cm^3$), with an optical window (measuring $8 \times 20$ $cm^2$). It can reach an ultimate vacuum pressure of $10^{-6}$ mbar and is equipped with a droplet injection unit, vacuum pump inlet, and nitrogen supply. The vacuum pressure is measured using a digital Pirani gauge (make: Ultra-High Vacuum, range: 999 to 0.001 mbar), while the gauge pressure is measured using a digital pressure sensor (make: Janatics Pneumatic, range: 0 to 10 bar). To achieve the inert conditions, the chamber is first evacuated using a vacuum pump up to the absolute pressure level of 0.018 mbar. The chamber is then purged with ultra-high pure nitrogen gas (Chemix gases, Nitrogen grade 6.0, Oxygen concentration $<1$ ppm) until the gauge pressure of 0.35 bar is achieved inside the chamber. This process is repeated for five cycles until the oxygen concentration inside the test chamber is reduced below 1 ppm level (estimated through theoretical means). Note that the nitrogen supply coming from the gas cylinder is further purified using oxygen and moisture traps installed between the cylinder exit and chamber inlet. After completing the purging cycles, the chamber is pressurized to 0.35 bar gauge pressure. This overpressure ensures that any leakage, if present, will occur from the inside to the outside of the chamber. The chamber is also tested to ensure that an absolute pressure of $>1$ mbar can be sustained for over 24 hours without the operation of the vacuum pump, which indicates minimal leakage is expected during experimental runs (each trial usually takes 20-30 minutes). These complex procedures are necessary to prevent the oxidation of galinstan during experimental runs. The atomization of Galinstan-air and water-air cases are also performed in the chamber with the same gauge pressure levels of 0.35 bar to maintain consistency. 

\section{Results and discussion}
\label{sec:Results and discussion}

\subsection{Global view of atomization dynamics}
\label{subsec:Global view of atomization dynamics}

\begin{figure}
\centering
\includegraphics[width=1\textwidth]{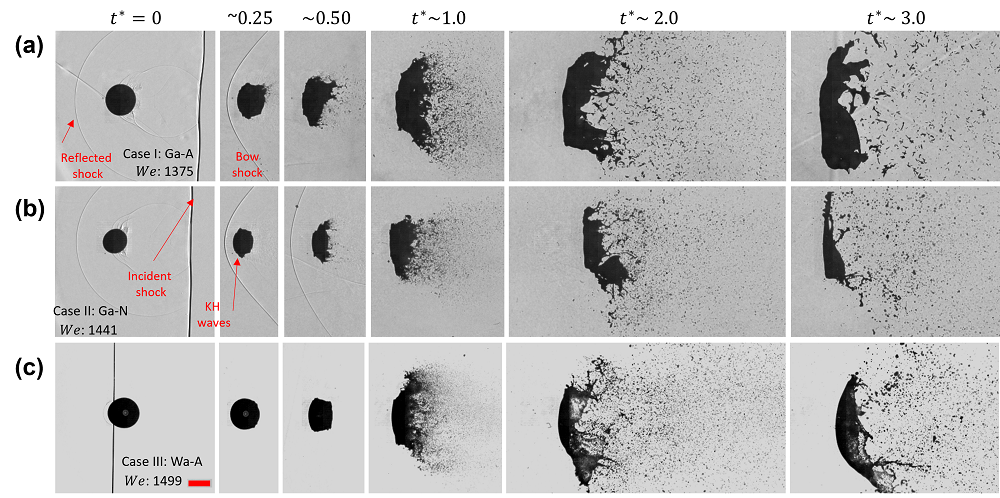}
\caption{Global view of atomization dynamics. (a) Case I: Galinstan-air test case (Ga-A) with $We$=1375, (b) Case II: Galinstan-nitrogen test case (Ga-N) with $We$=1441, (c) Case III: Water-air test case (Wa-A)  with $We$=1499. Scale bar equal 2 mm. See supplementary movie 1 for the video file.}
\label{Ch4:fig:global view}
\end{figure}

A global view of shock droplet interaction for three test cases is presented in figure \ref{Ch4:fig:global view}. The time sequence images are arranged from left to right, and absolute time is normalized with the inertial time scale $t_{in}$, where $t_{in}= D_o\sqrt{\frac{\rho_l}{2(P_o-P_1)}}$.  Each row of images represents three test cases, namely Galinstan-air, Galinstan-nitrogen, and water-air. The test cases are visualized at the same non-dimensional time scale ($t^*$) instant. The first column of images at $t^*=0$ represents the moment of shock droplet interaction. Figure \ref{Ch4:fig:global view} compares the three test cases at comparatively larger spatial and temporal scales, providing a global view comparison of the atomization dynamics.  See supplementary movie 1 for the video files.

\hspace{0.2cm} \indent In all three test cases, after the interaction of the incident shock wave, different shock structures, such as a reflected wave, Mach stem, etc., are formed, which instantaneously change the local flow and pressure conditions around the droplet (see figure \ref{Ch4:fig:global view} at $t^* = 0$). This wave dynamics is short-lived and is followed by the induced airflow interaction with the droplet \citep{sharma2021shock}. The surface tension of galinstan is higher than DI water; therefore, to maintain the same Weber number values for galinstan and water, the shock strength in the case of galinstan has to be increased significantly (see figure \ref{Ch4:fig:Weber number regime}b), resulting in the generation of shock-induced flow to be at supersonic conditions. The deflection of such supersonic gas flow from the droplet surface will create a bow shock (see figure \ref{Ch4:fig:global view}a,b at $t^* \sim 0.25$) in front of the droplet windward surface. Based on the shock standoff distance from the windward droplet surface,  induced gas flow Mach numbers are estimated to range between 1.1 to 1.3 \citep{starr1976shock}. No such bow shock formation is observed for a DI water droplet, as the induced airflow is essentially in the subsonic regime.

\hspace{0.2cm} \indent As time progresses, droplet deform along with a simultaneous formation of Kelvin Helmholtz waves on the windward surface, subsequently forming a thin liquid sheet at the droplet's pole from where stripping of secondary droplets begins ($t^* \sim 0.5$ to $3.0$) \citep{sharma2021shock}.  At each time instant, droplet morphology appears to be similar in all three test cases. Shear-induced entrainment (SIE) mode of droplet breakup is observed in all test cases studied in this work. Droplet deforms into a cupcake shape until $t^* \sim 0.25$ and stripping of droplets starts at $t^* \sim 0.5$. Further, a significant extent of sheet atomization is achieved up to $t^* \sim 1$, which is followed by the recurrent breakup.

\hspace{0.2cm} \indent A closer look at the images corresponding to $t^* > 1.0$ suggests differences in the fragment shapes of the Galinstan-air case when compared to the other two test cases. Flake-like fragments are formed during Galinstan atomization in air. Irregular-shaped fragments are visualized more clearly in figure \ref{Ch4:fig:atomization with increasing Weber number}. Such flake formation is attributed to the formation of an oxide layer on the fragmented liquid surface \citep{chen2018galinstan}, which inhibits further breakup and locks the liquid metal into irregular shape metal oxide shells. Similar observations for the three test cases are observed for other Weber number values tested in this work (see supplementary figures S2 and S3). It should be noted that the breakup period of a Galinstan droplet in the air environment is of the order of $\sim 100$ $\mu$s which itself is observed to be sufficient for the creation of a significant metal oxide layer on the surface of the secondary droplets, thereby inhibiting the atomization process. This suggests at least a microsecond duration  estimation  for the Galinstan oxidation time in ambient air,  much faster than the previously observed oxidation time scales based on the droplet impact experiments \citep{kim2013recovery}.  In contrast to the Galinstan-air test case, atomization dynamics of a Galinstan-nitrogen test case is essentially similar to that of DI water at comparable Weber number values, and spherical-shaped secondary droplets are formed in the two cases.

\subsection{Mechanism of liquid metal droplet breakup in SIE mode}
\label{subsec:Mechanism of Galinstan droplet breakup in SIE mode}

\begin{figure}
\centering
\includegraphics[width=1\textwidth]{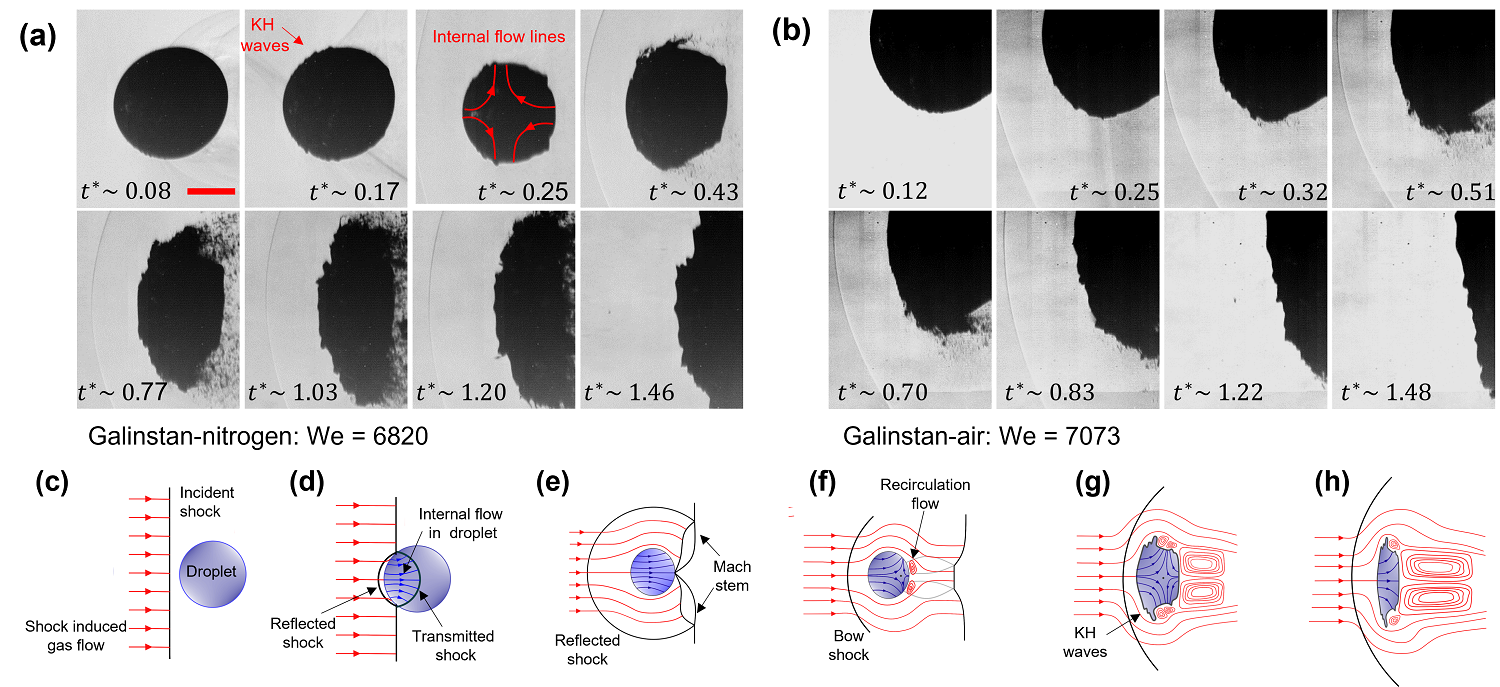}
\caption{Mechanism of Galinstan droplet breakup in SIE mode. (a) Zoomed in view of Galinstan droplet breakup in an inert environment at $We =6820$. (b) Zoomed in view of Galinstan droplet breakup in an air environment at $We =7073$.(c-h) Schematic diagram showing the mechanism of  SIE breakup during SIE mode. Scale bar equals 1mm.  See supplementary movie 2 for the video file.}
\label{Ch4:fig:mechanism of galinstan breakup}
\end{figure}

The previous subsection (\ref{subsec:Global view of atomization dynamics}) presented a global overview of atomization dynamics for three test fluids. This subsection compares the detailed mechanism of three test cases undergoing droplet breakup through SIE mode. High spatial and temporal resolution were employed to uncover the mechanism details. Figure \ref{Ch4:fig:mechanism of galinstan breakup}a,b provides a zoomed-in view of a Galinstan droplet atomization in nitrogen and air environment, respectively.  See supplementary movie 2 for the video file. Observation similar to the Galinstan-nitrogen test case is observed for the DI water-air test case (see supplementary figure S4). A schematic diagram illustrating different events during interaction is shown in figure \ref{Ch4:fig:mechanism of galinstan breakup}c-h. The wave structures depicted in the schematics, such as incident shock, reflected shock, Mach stem, etc., are based on the present visualisations and our previous publication \citep{sharma2021shock}. As flow dynamics could not be visualized through experimental means, evidence from several numerical simulations is utilized to construct flow fields \citep{Das_2020,Guan_2018,Sembian_2016}. 

\hspace{0.2 cm} \indent After interacting with the droplet's windward surface, the incident wave transforms into a reflected wave and a transmitted wave. The reflected wave is visualized on several occasions (see figure \ref{Ch4:fig:global view} a,b at $t^* \sim 0$), while the transmitted wave could not be visualized through shadowgraphy imaging. Based on the acoustic impedance of gas and liquid fluids, a shock wave is formed as a reflected wave \citep{henderson1989refraction}. For an incident shock wave, the transmitted wave is also a shock wave \citep{henderson1989refraction,Sembian_2016}. The transmitted wave induces an internal flow inside the droplet (see figure \ref{Ch4:fig:mechanism of galinstan breakup}d), while the reflected wave diverts the streamlines of shock-induced gas flow (see figure \ref{Ch4:fig:mechanism of galinstan breakup}e). Initially, a regular reflection is observed until the angle of incidence of the incident wave reduces below a critical value. This is followed by the transformation of regular reflection to Mach reflection. The formation of Mach stems can be visualized in figure \ref{Ch4:fig:global view} at $t^* =0$. The reflected shock and Mach stem formation (and their collision at later time instances, see figure \ref{Ch4:fig:mechanism of galinstan breakup}e) increases the local pressure around the droplet, resulting in the formation of high-pressure regions on the windward and leeward sides when compared to the droplet pole. It must be noted that the initial wave dynamics is short-lived ($O(10)$ $\mu$s) when compared to the droplet breakup time ($O(100)$ $\mu$s), therefore all these wave features have a minimal effect on the breakup mechanism. In later time instances, the breakup dynamics is primarily governed by the induced gas flow interaction with the droplet (see figure \ref{Ch4:fig:mechanism of galinstan breakup} a,b for $t^* > 0.1$ and figure \ref{Ch4:fig:mechanism of galinstan breakup} f-h). 

\hspace{0.2cm} \indent Induced flow separation occurs at the droplet's leeward surface, creating a jet flow towards the rear stagnation point (see figure \ref{Ch4:fig:mechanism of galinstan breakup}g-h) \citep{Guan_2018,Das_2020}. The formation of the stagnation point on the windward, as well as the leeward side of the droplet, increases the local pressure in these regions compared to the droplet pole, inducing an internal flow towards the droplet pole and resulting in the droplet deformation into a cupcake shape (see figure \ref{Ch4:fig:mechanism of galinstan breakup}a at $t^* \sim 0.43$).  On the other hand, the relative velocity between the gas phase and liquid phase
lead to KH waves-based surface instabilities on the droplet’s windward surface. Surface instabilities get entrained with the external gas flow and are transported along the droplet periphery, resulting in the formation of a liquid sheet at the equator region (see figure \ref{Ch4:fig:mechanism of galinstan breakup}a,b at $t^* = 0.4$ to $0.8$). The formation of KH waves on the droplet windward side is also observed in the previous works \citep{Theofanous2008,liu2018numerical,sharma2021shock,chandra2022shock}. 

\hspace{0.2cm} \indent The above discussion suggests the two mechanisms for the formation of liquid sheets at the droplet pole i.e. liquid transport through KH waves entrainment and droplet deformation. Fragmentation of this liquid sheet on the droplet pole results in the formation of secondary droplets. The temporal size distribution of the atomized droplets is influenced by these two mechanisms, as discussed in the following subsection.

\subsection{Atomization dynamics of liquid metal droplet at high Weber number}
\label{subsec:Atomization dynamics of Galinstan droplet at high Weber number}

\begin{figure}%
\centering
\includegraphics[width=1\textwidth]{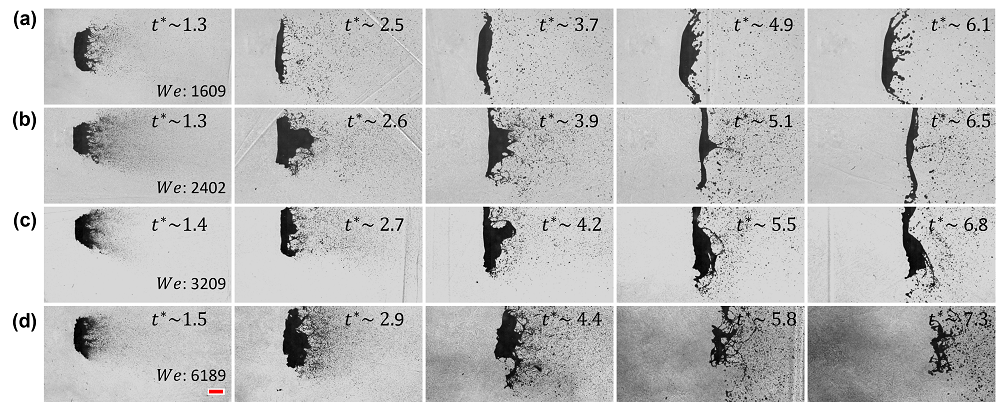}
\caption{Atomization dynamics of  Galinstan droplet in an inert environment, for different Weber number ($We$) values. (a) Galinstan-nitrogen $We$=1609, (b) Galinstan-nitrogen $We$=2402, (c) Galinstan-nitrogen $We$=3209, and (d) Galinstan-nitrogen $We$=6189. Scale bar equal 2 mm.}
\label{Ch4:fig:Ga N atomization with increasing Weber number}
\end{figure}

\begin{figure}%
\centering
\includegraphics[width=1\textwidth]{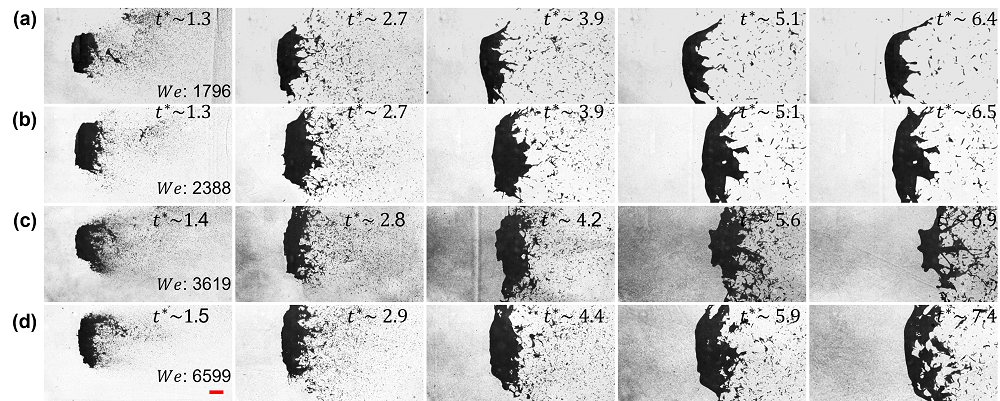}
\caption{Atomization dynamics of  Galinstan droplet in the air environment, for different Weber number ($We$) values. (a) Galinstan-air $We$=1796, (b) Galinstan-air $We$=2388, (c) Galinstan-air $We$=3619, and (d) Galinstan-air $We$=6599. Scale bar equal 2 mm.}
\label{Ch4:fig:atomization with increasing Weber number}
\end{figure}

 Figure \ref{Ch4:fig:Ga N atomization with increasing Weber number} and \ref{Ch4:fig:atomization with increasing Weber number} display time-series images (from left to right) of Galinstan droplet breakup in a nitrogen and air environment, respectively. Each row of images is organized in increasing order of Weber number values. According to the previous subsection, a liquid sheet is formed at the droplet's pole due to two mechanisms: liquid transportation based on Kelvin-Helmholtz (KH) waves and bulk flow resulting from droplet deformation. The first mechanism involves KH waves accumulating on the pole region, which produces a thinner sheet and smaller secondary droplets. The second mechanism involves liquid transport from the stagnation point to the pole region due to droplet deformation, resulting in bulk liquid transport and thus, a thicker liquid sheet. As a consequence, larger secondary droplets are produced due to this mechanism.
 
 \hspace{0.2cm} \indent During the initial stages of droplet breakup, sheet formation is predominantly influenced by the Kelvin-Helmholtz (KH) wave-based mechanism as droplet deformation is minimal (see figures \ref{Ch4:fig:mechanism of galinstan breakup}a,b at $t^* < 1$ and \ref{Ch4:fig:KH waves quantification}a,b ). Therefore, smaller secondary droplets are generated at this stage. However, in later time instances, liquid transport through droplet deformation takes over, resulting in the formation of larger secondary droplets. Additionally, multiple KH waves combine to form a thicker sheet, leading to the creation of  bigger secondary droplets (see figure \ref{Ch4:fig:mechanism of galinstan breakup}a at $t^* \sim 0.77$). This trend of increasing secondary droplet sizes with time is observed in all the test cases depicted in figure \ref{Ch4:fig:Ga N atomization with increasing Weber number} and \ref{Ch4:fig:atomization with increasing Weber number} and has also been quantitatively measured in our earlier study on aerobreakup of DI water droplets \citep{sharma2023depth}. Furthermore, as the Weber number value increases, the KH wave's wavelength decreases (explained later in \S\ \ref{subsec:Quantitative estimation of droplet deformation and KH waves magnitude}), and more KH waves are formed on the windward side. As a result, more droplet liquid is transported through the KH wave-based mechanism, leading to the generation of smaller fragments for higher Weber number values. It is observed that the droplet fragmentation process persists until the aerodynamic force on the droplet, which is related to the relative velocity between the gas phase and liquid phase, dominates the surface tension force. This condition is not sustained during the observation period for the lower Weber number values (see figure \ref{Ch4:fig:Ga N atomization with increasing Weber number} and \ref{Ch4:fig:atomization with increasing Weber number} a,b). However, at higher Weber numbers, complete fragmentation of the primary droplet occurs due to the higher aerodynamic forces acting on the droplet (see figure \ref{Ch4:fig:Ga N atomization with increasing Weber number}d and \ref{Ch4:fig:atomization with increasing Weber number}d).

\begin{figure}%
\centering
\includegraphics[width=1\textwidth]{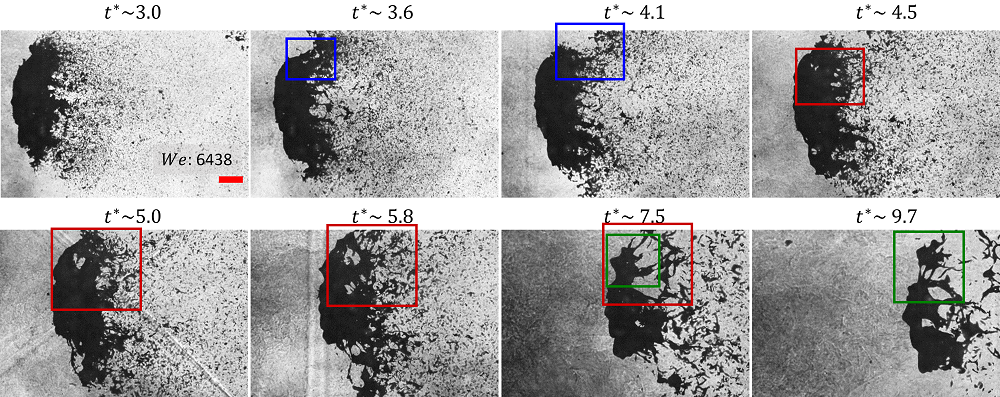}
\caption{ Recurrent breakup of Galinstan droplet in the air for $We = 6438$. The blue, red, and green rectangle indicates the first, second, and third breakup cycles. Scale bar equals 2 mm.}
\label{Ch4:fig:recurrent breakup}
\end{figure}

 \hspace{0.2 mm} \indent It is important to note that the droplet breakup during SIE mode is a repetitive process \citep{Dorschner_2020,sharma2021shock}, as illustrated in figure \ref{Ch4:fig:recurrent breakup}. During this process, the liquid sheet that forms at the droplet's pole is carried in the direction of the gas flow due to its entrainment with the external gas phase. As a result, the sheet is deflected and elongated in the flow direction, causing a continuous reduction in its thickness. This thin sheet is susceptible to instabilities, leading to the formation of holes (as shown in figure \ref{Ch4:fig:recurrent breakup}, at $t^* \sim 3.6$). Multiple holes are formed at various locations on the sheet, and they expand in all directions due to the surface tension force. Eventually, several holes collide with one another, resulting in the formation of ligaments. In the Galinstan-nitrogen test case, these ligaments undergo further pinch-off into secondary droplets, whereas, in the Galinstan air test case, the formed ligaments do not pinch off further due to significant surface oxidation at this time period, leading to the formation of irregular-shaped fragments (discussed in \S\ref{subsec:Phenomenological framework for secondary droplets morphology}). As discussed earlier, SIE breakup follows a repetitive cycle. First, a sheet is formed, followed by hole formation, ligament generation, and droplet pinch-off. This process is illustrated by the blue rectangle in figure \ref{Ch4:fig:recurrent breakup}. After one cycle, the same process repeats until the droplet disintegrates completely or until the restoring surface tension forces overpower the aerodynamic forces. This recurrent breakup of droplet is represented by the red and green rectangles in figure \ref{Ch4:fig:recurrent breakup}.

\subsection{Quantitative comparison for three test cases}
\label{subsec:Quantitative estimation of droplet deformation and KH waves magnitude}

\begin{figure}%
\centering
\includegraphics[width=1\textwidth]{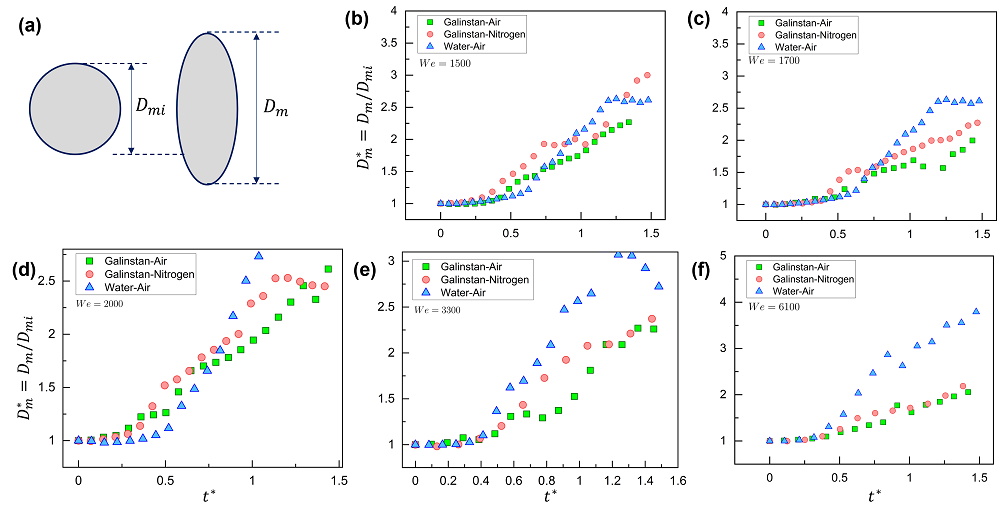}
\caption{Cross stream  droplet deformation, $D_{mi}$ and $D_{m}$ represent initial and instantaneous cross stream diameters. (a) Schematic diagram showing cross-stream deformation.  (b-f) Cross stream deformation with non-dimensional time for $We \sim 1500, \hspace{1mm}1700,\hspace{1mm}2000,\hspace{1mm}3300$ and $6100$}
\label{Ch4:fig:deformation quantification}
\end{figure}

The previous subsections (\S\ \ref{subsec:Global view of atomization dynamics}, \ref{subsec:Mechanism of Galinstan droplet breakup in SIE mode}, \ref{subsec:Atomization dynamics of Galinstan droplet at high Weber number}) provided a qualitative comparison of the atomization dynamics for three test cases. This subsection aims to make a quantitative comparison between them. Figure \ref{Ch4:fig:deformation quantification} shows the non-dimensionalized cross-stream diameter ($D_m ^*$) of the droplet plotted against non-dimensional time ($t^*$). To measure the instantaneous cross-stream diameter ($D_m$), an ellipse profile is fitted onto the deformed droplet using \textit{ImageJ} software. The results indicate a similar deformation trend  for all three test cases when $t^* < 0.75$, but some discrepancies exist for $t^* > 0.75$. As seen in Figure \ref{Ch4:fig:global view}, up to $t^* < 0.75$, the droplet has undergone deformation and KH wave formation, which is similar in all three test cases. After $t^* > 0.75$, droplet atomization begins, which makes it difficult to measure the cross-stream diameter accurately. This is because the cross-stream length of the liquid sheet can also be detected as the cross-stream diameter based on the measurement approach followed here. However, it is still essential to present the measurement of cross-stream deformation beyond $t^* > 0.75$ because it also indicates the cross-stream spread of the atomizing liquid sheet. This spread becomes more random during later time periods due to variations in the atomization behavior of a deformed droplet sheet on a case-to-case basis.

\begin{figure}
\centering
\includegraphics[width=1\textwidth]{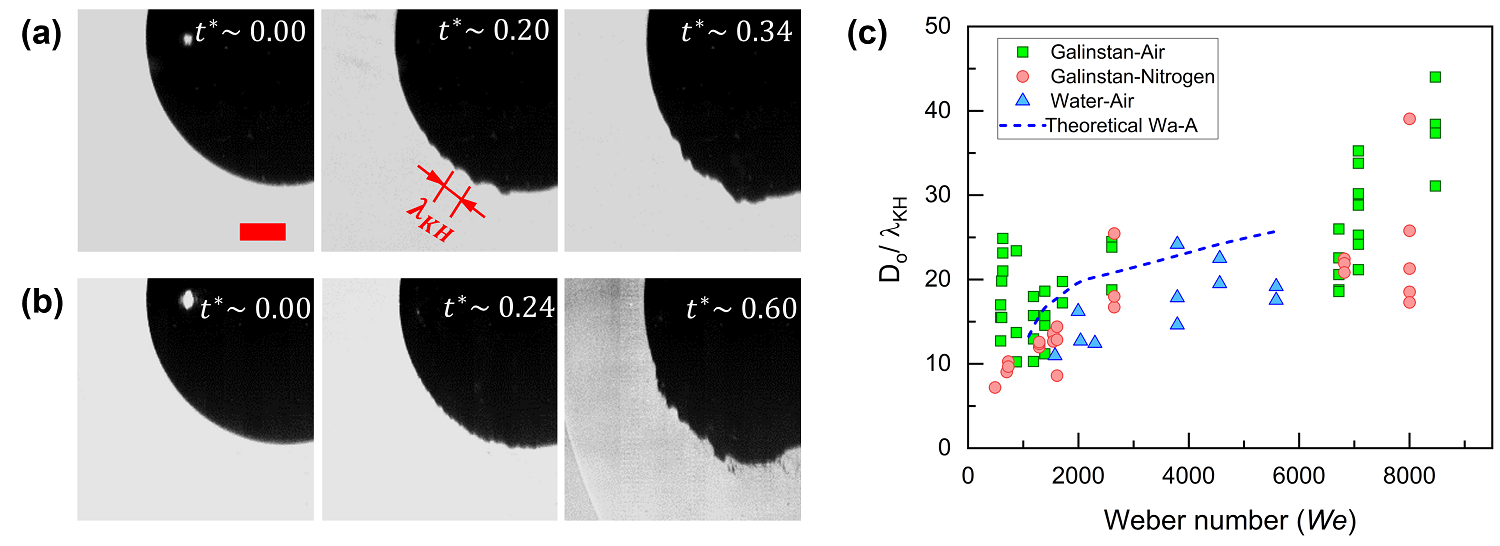}
\caption{Kelvin-Helmholtz instability waves formation on the windward surface of the droplet. (a,b)  Zoomed-in images showing the formation and evolution of KH waves on the surface of Galinstan droplet in air environment at $We =1186$ and $7073$, respectively. Scale bar equals 500 $\mu$m. (c) Variation of KH waves wavelength with different Weber number values. A blue dashed line shows the predicted wavelength values from equation \ref{eq:KH wave dispersion relation} for the water-air test case at different $We$ values.  See supplementary movie 3 for the video file.}
\label{Ch4:fig:KH waves quantification}
\end{figure}

\hspace{0.2cm} \indent Figure \ref{Ch4:fig:KH waves quantification} (also supplementary movie 3) presents the comparison of  experimentally measured of Kelvin-Helmholtz (KH) waves wavelength for three test cases using high spatial and temporal resolution measurements. Figure \ref{Ch4:fig:KH waves quantification}a at $t^* \sim 0.20$ displays a sample case of KH wave measurement. Figure \ref{Ch4:fig:KH waves quantification}a,b shows a zoomed-in image of KH wave formation for the Galinstan-air test case at $We = 1186$ and $7073$, respectively. The measured values of non-dimensionalized KH waves wavelength ($D_o/\lambda_{KH}$) for three test cases at different Weber number values are shown in Figure \ref{Ch4:fig:KH waves quantification}c. A theoretical framework of KH wave-based linear stability analysis of non-zero vorticity thickness for an incompressible fluid flow was provided in \cite{villermaux1998mixing,marmottant2004spray,padrino2006shear} also used in \cite{chandra2022shock,sharma2021shock}. A similar analysis is used to estimate the KH wavelength for the water-air test case. The dispersion relation can be written as:

\begin{equation}
    \label{eq:KH wave dispersion relation}
    e^{-2 \eta} = [1+(\Omega - \eta)] \left[\frac{ \Phi+(\Omega+\eta) \{ 2\Tilde{\rho}-(1+\Tilde{\rho})(\Omega + \eta)-(1+\Tilde{\mu})\beta \eta^2 \} }{\Phi+(\Omega+\eta) \{ 2\Tilde{\rho}-(1-\Tilde{\rho})(\Omega + \eta)-(1-\Tilde{\mu})\beta \eta^2 \}} \right]
\end{equation}
here $\Omega = -2\omega_{KH} \delta / V_i$ is dimensionless growth rate and $\delta$ is the thickness of vorticity layer. $\eta =k_{KH} \delta$ is a dimensionless wave number and $k_{KH}=2 \pi / \lambda_{KH} $ is the wavenumber. $\Tilde{\mu} = \mu_g/\mu_l$ and $\Tilde{\rho} = \rho_g/\rho_l$ are non-dimensionalized viscosity and density ratios, where subscript $i \And l $ represents gas and liquid phase respectively. $\beta$ and $\Phi$ are related to vorticity thickness based Weber number, and their definition can be seen from \cite{chandra2022shock}.
Equation \ref{eq:KH wave dispersion relation} is solved to obtain the wavelength of the fastest growing wavelength ($\lambda_{KH}$). The variation of this wavelength for different values of Weber number is shown in figure \ref{Ch4:fig:KH waves quantification}c. A good prediction is provided by equation \ref{eq:KH wave dispersion relation} (shown by blue dashed line) with experimental data for the water-air test case. It is not possible to extend the same mathematical approach for galinstan test cases because the induced gas flow condition in the present studies is in a lower supersonic regime (i.e., flow Mach number $<$ 1.2), and reducing governing mass and momentum equations into a linearized form is not possible in this flow regime \citep{anderson1990modern}. The wavelength of KH waves decreases with an increase in the $We$ values. This observation is consistent with previous measurements obtained for Newtonian \citep{sharma2021shock} and viscoelastic fluids \citep{chandra2022shock}. Further, another aspect covering the criteria for SIE mode breakup, which was discussed in \cite{sharma2021shock,Theofanous_2012}, is validated in the present work. It was demonstrated that the SIE mode only exists if $D_o/ \lambda_{KH} >> 1$, and this condition is validated in the present work for liquid metal droplets. \\
\indent\hspace{0.2cm} The discussion from \S\ \ref{subsec:Global view of atomization dynamics}, \ref{subsec:Mechanism of Galinstan droplet breakup in SIE mode}, and \ref{subsec:Atomization dynamics of Galinstan droplet at high Weber number} suggests  that the mode and mechanism of droplet breakup are qualitatively and quantitatively similar for all three test cases, even at the time and length scales of breakup induction and KH wave formation, respectively. The above discussion provides important insights that the governing physics of liquid metal atomization can be explored through conventional fluids, and results can be translated to the atomization of liquid metals provided similar Weber number values and environmental conditions are maintained. However, the results of an oxidizing test case can only be compared regarding the modes and mechanisms of the secondary droplet formation. Their morphology differs significantly from the inert cases,  as discussed in the following subsection.

\subsection{Phenomenological framework for secondary droplets morphology}
\label{subsec:Phenomenological framework for secondary droplets morphology}

\begin{figure}%
\centering
\includegraphics[width=1\textwidth]{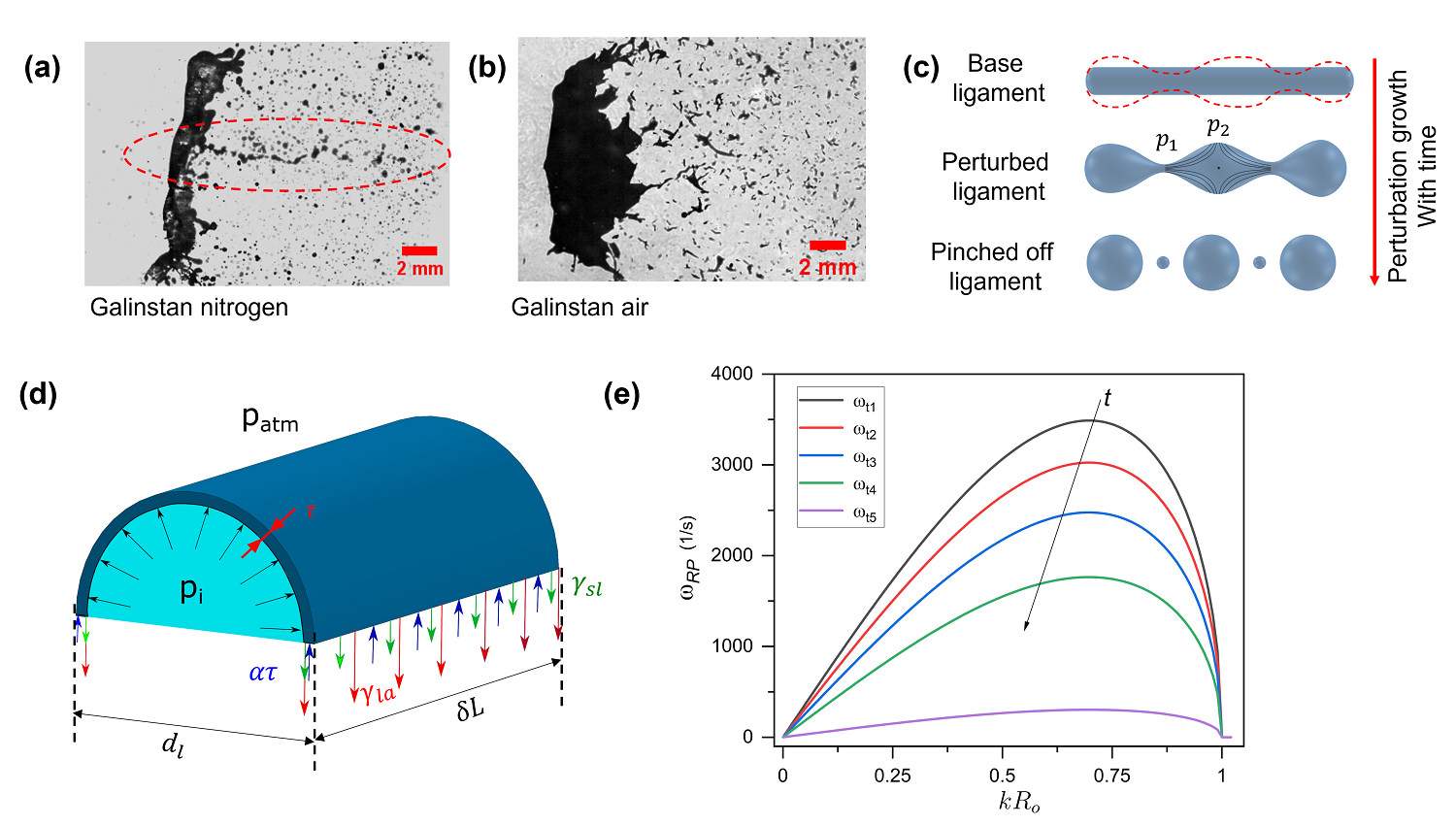}
\caption{Phenomenological framework for secondary droplets morphology. (a) Atomization of Galisntan  ligament through  Plateau Rayleigh (PR) instability in a nitrogen environment. (b) Stable ligament formation during Galinstan breakup in air atmosphere. (c) Schematic diagram showing instability growth through PR instability. (d) Schematic showing different interfacial forces acting on the Galinstan liquid ligament  when exposed to air environment. (e) The growth rate of PR instability at different time instances.}
\label{Ch4:fig:capillary breakup criteria}
\end{figure}

In the previous discussion, we compared the qualitative and quantitative aspects of three test cases concerning breakup modes, atomization mechanism, deformation characteristics, and KH wave wavelength magnitude. In this section, we will focus on the differences in these cases based on the breakup morphologies and provide a theoretical framework to predict each morphology type. As we discussed in \S\ref{subsec:Global view of atomization dynamics}, the breakup morphology of the DI water case and Galinstan nitrogen case is similar, where spherical-shaped secondary droplets are formed. In contrast, the Galinstan-Air test case results in flake-like fragments due to the rapid oxidization of the fragmenting ligaments. Figure \ref{Ch4:fig:capillary breakup criteria}a,b depicts the ligament breakup for Galinstan-nitrogen and Galinstan-air test cases, respectively. It should be noted that a liquid ligament under small perturbation is prone to surface instabilities, which can grow over time due to surface tension forces, leading to the pinch of ligaments into numerous secondary droplets. This ligament breakup mechanism is well-known as Plateau-Rayleigh (PR) instabilities \citep{drazin2004hydrodynamic,sharma2021secondary} and is schematically represented in figure \ref{Ch4:fig:capillary breakup criteria}c. When a small perturbation occurs on the base ligament, it results in differential pressures $p_1$ and $p_2$ at two locations. The growth rate ($\omega$) of PR instability can be calculated using the following expression \citep{drazin2004hydrodynamic}:

\begin{equation}
    \label{eq:PR instability}
    \omega^2 = \frac{\sigma}{\rho_l R_l^3}kR_l \frac{I_1(kR_l)}{I_0(kR_l)}(1-k^2R_l^2)
\end{equation}

here $R_l$ represents the initial radius of the ligament, $I_1$ and $I_0$ are first and zero-order modified Bessel functions of the first kind, respectively, and $k$ is the wave number. The time scale for ligament rupture can be determined using equation \ref{eq:PR instability}. Specifically, the time scale is given by $t_{c} \sim \frac{1}{\omega} \sim \sqrt{\frac{\rho_l R_l^3}{\sigma}}$. For the Galinstan-nitrogen cases, this time scale is around $O(100)$ $\mu$s for ligaments with a diameter of $O(100)$ $\mu$m, which aligns with experimental findings.

\hspace{0.2cm} \indent In the Galinstan-air test case, a thin elastic metal oxide layer forms on the ligament surface due to exposure to an oxidizing environment. Based on the present experimental observation, the rate of oxide layer formation is inferred to be much faster than the experimental timescales($O(100)$ $\mu$s). This suggests an instantaneous growth rate of the oxide layer (compared to experimental time scales). It has been shown in the literature that the surface characteristics of fully oxidized gallium-based alloy are mainly influenced by the formed oxide layer \citep{dickey2008eutectic,handschuh2021surface}. The elasticity of the oxide layer counteracts the effective interfacial tension (due to solid-liquid and liquid-air interface), resulting in a negligible net normal interfacial force acting on the ligaments. Consequently, as per PR instability analysis, stable ligaments are formed, as in the case of the Galinstan-air test case.

\hspace{0.2cm} \indent The oxidation rate of liquid metals can be controlled by modifying the oxygen concentration in the environment. In the current study, Galinstan was tested under two extreme conditions: ambient air, where there is sufficient oxygen supply, and an inert atmosphere with an extremely low oxygen concentration. This results in two different types of breakup morphologies (spherical droplets and flakes-like fragments). The stability of the ligaments and the type of breakup morphology observed are dependent on the oxidation rate of the liquid metal and the resulting properties of the oxide layer formed on the ligament surface. If the oxygen concentration is in between the above two extremes, the oxidation rate of liquid metal is not instantaneous. In such cases, the size of fragmenting ligaments determines the shape of secondary droplets. A typical example of such a ligament is depicted in figure \ref{Ch4:fig:capillary breakup criteria}d. The pinch-off of these ligaments is influenced by interfacial forces (such as $\sigma_{la}$ for the liquid Galinstan-air interface and $\sigma_{sl}$ for the solid gallium oxide-liquid Galinstan interface), and a thin metal oxide layer which provides an elastic resistance to the pinch-off process, resulting in circumferential compressive stress. As a result, the normal force balance on the interface is given by the following expression.

 \begin{equation}
     \label{eq: normal force 1}
     (p_i - p_{atm})\delta L d_l = 2[\phi(t) \sigma_{sl} \delta L + \psi(t) \sigma_{la} \delta L- \alpha \tau(t)]  \\
\end{equation}

here $p_i$ and $p_{atm}$ represent the internal and ambient pressure of the ligament, respectively. $\delta L$ represents the infinitesimal element of the ligament, and $d_l$ represents the diameter of the ligament. The weight factors $\phi(t)$ and $\psi(t)$ are time-dependent and control the effective interfacial tension of the solid-liquid and liquid-air interfaces, respectively. $\alpha$ represents the yield stress of the oxide layer, and $\tau(t)$ represents the time-varying thickness of the oxide layer. Initially, $\tau(t)$ and $\phi(t)$ will be zero, and $\psi(t)$ will be 1 since there will be no oxide layer at the start of the process ($t = 0$). The magnitudes of these factors will change over time based on the thickness of the oxide layer. The equation \ref{eq: normal force 1} can be further simplified as follows:

 \begin{equation}
     \label{eq: normal force 2}
     (p_i - p_{atm}) d_l = 2[\sigma_{effective}- \alpha bt]=2F_i (t)  \\
\end{equation}

here, $\sigma_{effective}$ is the effective interfacial tension and $b$ is the growth rate of oxide layer. $F_i$ represents the interfacial force per unit ligament length, which is equivalent to liquid-air interfacial tension when the oxide layer is not present. After following the linear stability for PR instability, using equation \ref{eq: normal force 2} one can get the growth rate of instability as:

\begin{equation}
    \label{eq:PR instability 2}
    \omega^2 = \frac{F_i}{\rho_l R_l^3}kR_l \frac{I_1(kR_l)}{I_0(kR_l)}(1-k^2R_l^2)
\end{equation}

The growth rate of instability ($\omega$) depends on $F_i$, which decreases with time. Consequently, the growth rate of instability is expected to decrease over time, as shown in figure \ref{Ch4:fig:capillary breakup criteria}e. If the pinch-off of a ligament occurs before the instability growth rate reaches zero, then we can expect the formation of spherical droplets, as observed in the Galinstan-nitrogen test case. However, if this criterion is not fulfilled, as in the case of the Galinstan-air test case, the formed ligaments do not undergo further atomization into secondary droplets. The time scale of $F_i \to 0$ is given by $t_o = \frac{\sigma_{effective}}{\alpha b}$ from equation \ref{eq: normal force 1}. For a typical ligament size of $O(100)$ $\mu$m we can consider two extreme situations to determine the possible oxidation rates of the Galinstan. The capillary breakup time scale for this ligament size is $O(100)$ $\mu$s. For the first instance, if we consider the instantaneous oxidation of the ligament, i.e. assuming time scale of oxide layer formation as $O(1)$ $\mu$s  and $\sigma_{effective} =\sigma_{sl}$, the oxidation rate ($b$) turned out to be of $O(1)$ $mm/s$. Here an estimation of $\alpha$ is taken from \cite{dickey2008eutectic} using oxide layer thickness estimates from \cite{scharmann2004viscosity,jia2019liquid,bilodeau2017liquid}. For the second instance, if the oxidation rate is very slow, as in the case of an inert atmosphere, $\sigma_{effective} =\sigma_{la}$ and the oxidation rate is estimated as $O(0.1)$ $\mu$m/s provided complete oxidation is assumed to occur at $O(10)$ ms.  Comparing $t_o$ with the time scale of capillary breakup $(t_c)$ gives an essential criterion for the ligament size that will undergo pinch-off into droplets for a given oxidation rate as:

\begin{equation}
    \label{eq: ligament radius}
    R_l << \left(\frac{\sigma_{effective} ^2 F_i}{\rho_l \alpha^2 b^2} \right)^{1/3} 
\end{equation}

The above-proposed framework provides a phenomenological explanation for the complete pinch-off of Galinstan ligaments in nitrogen environments and flake formation in air environments. The presented framework fits well for the two extreme oxidation rates considered in this work. However, verifying this framework for intermediate oxidation rate could not be obtained because estimation for the Galinstan oxidation rate is an onerous task. As per the author's knowledge, there is no existing literature available for such an estimation. The oxidation rate models provided for other liquid metals \citep{zhang2010high} cannot be translated here because of the very short time scale of the phenomenon in the present case. This difficulty arises due to the short time scale of the oxidation phenomenon (around $100$ $\mu$s) and the small length scales involved (around $O(1)$ nm). Therefore, currently, it is not possible to achieve quantitative validation of the proposed model for intermediate oxidation rates. Nevertheless, if a robust estimation of the Galinstan oxidation rate becomes available in the future, it could be used to feed values into the proposed framework, and quantitative validation of the model could be achieved.

\section{Conclusions}
\label{sec:Conclusions}
The aerobreakup of a liquid metal droplet at a high Weber number ($We \sim 400 - 8000$) regime is investigated. Galinstan is taken as a test fluid, and multi-scale spatial and temporal investigations are carried out for three test environments: oxidizing (Galintan-air), inert(Galinstan-nitrogen), and conventional fluids (water-air). Using Galinstan-nitrogen as a test case provides an excellent base for comparing liquid metal atomization results with conventional fluids, an approach considered for the first time in this work. A global view of atomization dynamics reveals similar qualitative and quantitative features between Galinstan-nitrogen and water-air test cases. This indicates that the results of conventional fluids can be translated to liquid metal droplets provided similar Weber numbers and environmental conditions are maintained. On the other hand, the Galinstan air test case showed a difference in the morphology of fragmenting secondary droplets due to the rapid oxidation of liquid metal, while the mode and mechanism of atomization remain similar to other cases. 

\hspace{0.2 cm} \indent For the considered $We$ values, all test fluids show a similar breakup mode and mechanism at multiple length scales and time scales explored in this work. Shear-Induced Entrainment (SIE) breakup mode is observed in all cases. After interacting with the droplet surface, the incident shock wave transforms into several wave structures which influence the local flow and pressure condition around the droplet for a short duration. However, the later-stage breakup mechanism is primarily governed by shock-induced gas flow interaction with the droplet. SIE mode is characterized by the fragmentation of a thin liquid sheet on the droplet pole, forming secondary droplets. The two mechanisms identified for forming the liquid sheet are liquid transport through KH wave entrainment and droplet deformation.  These two mechanisms influence the temporal size distribution of the atomized droplets, where KH waves/ deformation-based mechanisms lead to smaller/bigger secondary droplets, respectively. The monotonous increase in the secondary droplet sizes is explained based on the relative dominance of two mechanisms. The SIE mode of droplet breakup is a repetitive process involving sheet formation, hole formation, ligament generation, and droplet pinch-off. This process repeats until the droplet disintegrates completely or until surface tension forces overpower aerodynamic forces.

Quantitative similarities for the three test cases are also provided regarding the temporal evolution of non-dimensionalized cross-stream diameter and experimental measurement of Kelvin-Helmholtz (KH) surface waves. The cross-stream deformation is found to be similar for all test fluids until the beginning of droplet atomization. In all test cases, the wavelength of KH waves decreases with an increase in the Weber number values, consistent with previous measurements for Newtonian and viscoelastic fluids. Further, the criteria for SIE mode breakup (i.e. SIE mode only exists if $D_o/\lambda_{KH} >> 1$) is validated in the present work for liquid metal droplets.

\hspace{0.2cm} \indent The  morphology of DI water and Galinstan nitrogen secondary droplets are similar, where spherical-shaped secondary droplets are formed. However, in the third case (Galinstan-air), flake-like fragments are produced due to the rapid oxidation of the fragmenting ligaments, which inhibits further breakup and locks the liquid metal into irregular shape metal oxide shells. The Plateau-Rayleigh (PR) instability is responsible for the pinch-off of the ligaments into secondary droplets in the first two cases. A phenomenological framework is also provided, which explains the flakes-like secondary droplets formation in the Galinstan air test case. The postulated framework depends on the oxidation rate of the liquid metal and the properties of the oxide layer formed on the ligament surface. The oxidation rate of liquid metals can be controlled by modifying the oxygen concentration in the environment. When the oxidation rate is between the two extreme conditions (no oxidation and rapid oxidation), the size of fragmenting ligaments determines the shape of secondary droplets.

\section*{Acknowledgments}
The authors acknowledge support from IGSTC (Indo–German Science and Technology Center) through project no. SP/IGSTC-18-0003. N.K.C. acknowledges support from the Prime Ministers Research Fellowship (PMRF). The authors gratefully acknowledge Professor C. Tropea (TU Darmstadt, Germany), Professor S. Chakraborty (IIT KGP, India), 
 Professor I. Roisman (TU Darmstadt, Germany), SMS group (Germany), Tata steel (India) and IGSTC PPAM group for their insights and discussions during the development of the present work.

\section*{Supplementary movies}\label{sec:sup_vid}
Supplementary figures and movie files are also provided as supporting material.\\
\textbf{Movie 1}: Global view of atomisation.\\ 
\textbf{Movie 2}: Mechanism of Galinstan droplet breakup.\\
\textbf{Movie 3}: Kelvin-Helmholtz instability waves visualisation.\\

\section*{Declaration of Interests}

The authors report no conflict of interest.

\section*{Author ORCIDs.}
Shubham Sharma \hspace{2mm}https://orcid.org/0000-0002-8704-887X;\\
Navin Kumar Chandra \hspace{2mm}https://orcid.org/0000-0002-1625-748X;\\
Aloke Kumar \hspace{2mm}https://orcid.org/0000-0002-7797-8336.\\
Saptarshi Basu \hspace{2mm}https://orcid.org/0000-0002-9652-9966;\\

\bibliographystyle{jfm}
\bibliography{main.bib}
\end{document}